\documentclass[aps,
footinbib,
superscriptaddress,
showpacs,twocolumn,
pra]{revtex4-1}
\newcommand{\expec}[1]{\langle #1\rangle}

\usepackage{amsmath}
\usepackage{amsfonts}  
\usepackage{graphicx}
\usepackage{float}
\usepackage{array}
\usepackage{hyperref}
\usepackage{lipsum}
\usepackage[usenames]{color}
\usepackage{bbm}

\usepackage{braket}

\begin{document}

\title{
Geometric representation of spin correlations and applications to ultracold systems}

\author{Rick Mukherjee}
\affiliation{Department of Physics and Astronomy, Rice University, Houston, Texas 77005, USA}
\affiliation{Rice Center for Quantum Materials, Rice University, Houston, Texas 77005, USA}

\author{Anthony E. Mirasola}
\affiliation{Department of Physics and Astronomy, Rice University, Houston, Texas 77005, USA}
\affiliation{Rice Center for Quantum Materials, Rice University, Houston, Texas 77005, USA}

\author{Jacob Hollingsworth}
\affiliation{Department of Physics and Astronomy, Rice University, Houston, Texas 77005, USA}
\affiliation{Rice Center for Quantum Materials, Rice University, Houston, Texas 77005, USA}

\author{Ian G. White}
\affiliation{Department of Physics and Astronomy, Rice University, Houston, Texas 77005, USA}
\affiliation{Rice Center for Quantum Materials, Rice University, Houston, Texas 77005, USA}

\author{Kaden R. A. Hazzard}
\affiliation{Department of Physics and Astronomy, Rice University, Houston, Texas 77005, USA}
\affiliation{Rice Center for Quantum Materials, Rice University, Houston, Texas 77005, USA}

\begin{abstract}
We provide a  one-to-one map between the spin correlations and certain three-dimensional shapes, analogous to the map between single spins and Bloch vectors, and demonstrate its utility. Much as one can reason geometrically about dynamics using a Bloch vector -- e.g. a magnetic field causes it to precess and dissipation causes it to shrink --  one can reason similarly about the shapes we use to visualize correlations. This visualization demonstrates its usefulness by unveiling the hidden structure in the correlations. For example, seemingly complex correlation dynamics can be described as simple motions of the shapes. We demonstrate the simplicity of the dynamics, which is obscured in conventional analyses, by analyzing several physical systems of relevance to cold atoms.
\end{abstract}

\maketitle

\section{Introduction}

Correlations play an important role in all branches of science. They determine the distribution of galaxies in cosmology \cite{Maddox}, reveal complex structure of molecules and proteins in chemistry and biology \cite{Keeler,Nicolas},  are invaluable in quantum sensing and computing in the form of quantum entanglement \cite{Horodecki1995, Bruning, Vicente, DeVicente2, Linden, Laraoui,Strobel424}, and are used to test fundamental predictions of QCD about entangled quark pairs \cite{Aaltonen, Mahlon}. In particular, correlations are of fundamental interest in many-body physics as they characterize phases in condensed matter \cite{Sachdev, Fradkin, Stevenson} and in ultracold matter \cite{Navon, cooper,Pielawa,Carr, Sanchez, Trotzky, Endres,Schauss1455}.

\begin{figure}[h!]
	\centering
	\includegraphics[width=8.2cm]{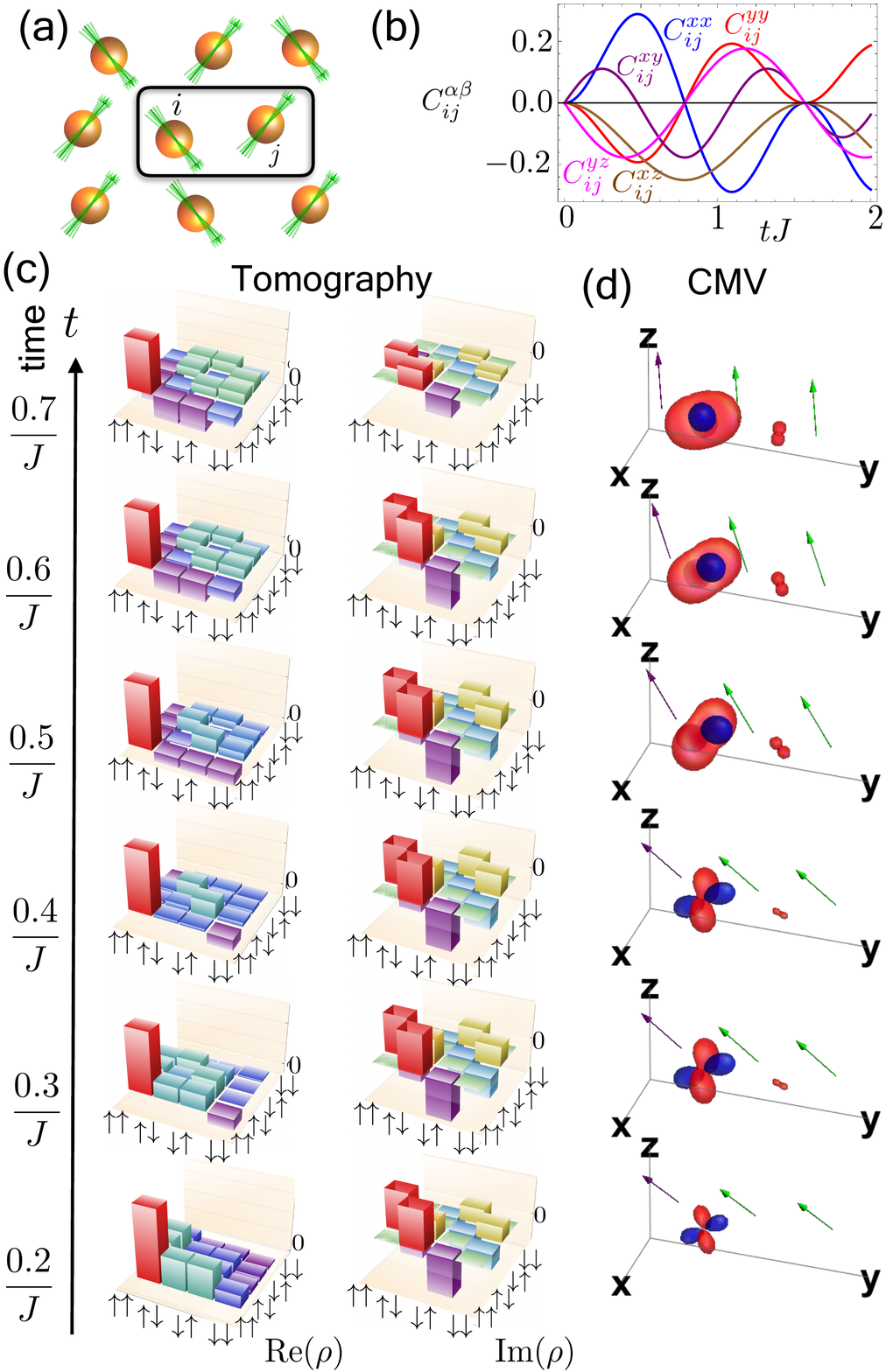}
	\caption{Unveiling of hidden structure in correlations by our visualization method. (a) Correlations in a many-body system of spin-$1/2$s. Green arrows are the Bloch vectors of each spin. (b) Complicated evolution of spin correlation components $C_{ij}^{\alpha\beta}$ versus time with $\alpha, \beta=x,y,z$ and $i,j$ index the spins. (This example is calculated for Ising dynamics, discussed in Sec.~\ref{Ising1}) (c) Tomographic representation of the two-spin density matrix for the same dynamics, showing $\braket{ab|\rho|cd}$ for $\{a,b,c,d\}\in \{\uparrow,\downarrow\}$. (d) Our visualization method (correlation matrix visualization = CMV) applied to the same dynamics reveals that the complicated components hide a very simple dynamics: a ``clover" shape grows in a plane perpendicular to the Bloch vector and rotates.}
	\label{setup}
\end{figure}

However even the simplest correlations in many-body systems have considerable complexity. For example, consider correlations between two spin-1/2s, as illustrated in Fig.~\ref{setup}(a). The correlations between each pair of spins $i$ and $j$ are described by $\expec{\sigma_i^\alpha \sigma_j^\beta}$ with $\alpha,\beta\in \{x,y,z\}$, and therefore require nine components to specify. Furthermore, as Fig.~\ref{setup}(b) illustrates, the behavior of these components can be complicated and seemingly structureless. Similarly, other standard visualizations such as density matrix tomography plots [Fig.~\ref{setup}(c)] fail to reveal any obvious structure. 

In this paper, we provide a method that encodes all of the correlation components holistically in a geometric object.  We demonstrate that seemingly complicated, structureless dynamics is in fact simple motions of these geometric shapes for several examples of many-body systems. For example, our visualization in Fig.~\ref{setup}(d) reveals the superficially complicated dynamics of Fig.~\ref{setup}(b-c) to be a  simple growth and rotation of an object in the shape of a clover. Our examples are drawn from recent experiments in ultracold matter, including lattice fermions \cite{Vekua, Dao}, Rydberg atoms \cite{Zeiher, Weimer1, Labuhn, Mukherjee}, molecules \cite{Alexey, Manmana, Yan}, and trapped ions\cite{Islam583,Kim, Blatt,britton:engineered_2012, Bohnet:quantum_2016}, but the visualization techniques are completely general.

Our work is not the first to consider geometric visualizations of spin correlations, but builds on the useful tools in Refs.~\cite{Feynman,Kimura,Byrd, Jakobczyk,Tilma, Bertlmann,Giraud,Jevtic, Dunkl,Kurzynski,SORENSEN, HALSTEAD,Donne,Philp,Merkel,Dowling, Harland,Gamel,Garon,Glaserapp}. We provide some new methods (see Sec.~III) to interpret the shapes, such as how to read off correlations from the shape.  Nevertheless, our construction is in essence equivalent to that in Refs.~\cite{Dowling, Harland,Gamel,Garon}. However, these prior studies considered quantum states chosen at random or for illustrative purposes. In contrast, we apply these tools to two-particle correlations in real many-body systems. In doing so, we show that not only are the shapes a compact way to summarize the spin components, but also that physical phenomena which appear complicated and mysterious when considering their components are simple, easy-to-describe motions when visualized geometrically.

This article first presents, in Sec.~\ref{recipe}, the prescription for generating the visualizations. Sec.~\ref{read} discusses how to interpret these correlations, whose size and color convey the magnitude and sign of the correlations. It also surveys the variety of shapes that can occur and their significance, and shows examples for prototypical quantum states. Sec.~\ref{real} applies these techniques to visualize the correlations of several systems -- both in nonequilibrium and equilibrium -- that are relevant to ultracold matter. It shows that in all cases the visualizations provide an especially simple representation of the behavior. Although we focus on two-particle correlations that are symmetric, our method can be generalized to include more spins and allow asymmetric correlations as briefly explained in Appendices~\ref{App0} and \ref{App2}.

\begin{figure*}[t]
	\centering
	\includegraphics[scale=0.55]{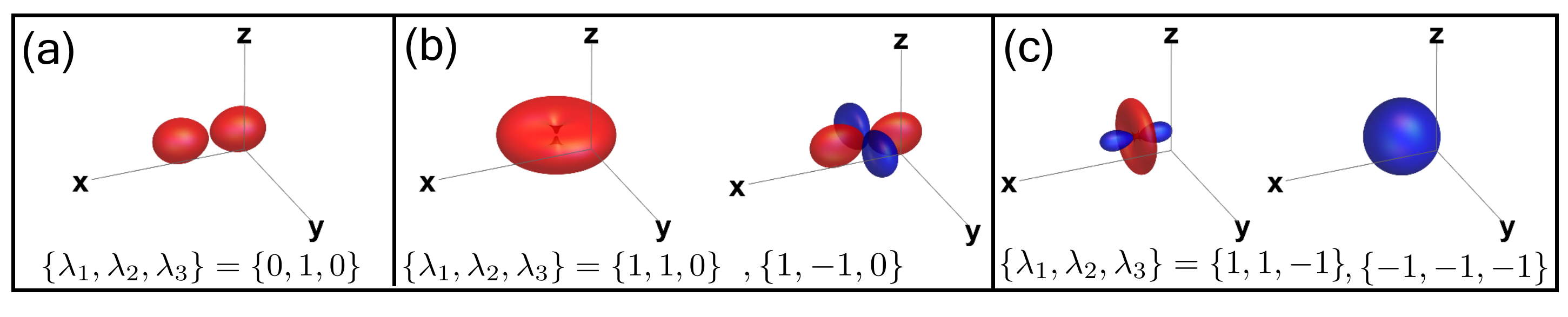}
	\caption{CMV shapes: The shape of the CMV is determined by the set of non-zero eigenvalues $\{\lambda_1,\lambda_2,\lambda_3\}$ of the correlation matrix $C_{ij}$ and their relative signs.  Positive (negative) correlations are depicted as red (blue) CMV surfaces. (a) CMVs with a single non-zero eigenvalue (rank-1 CMVs) are dumbbells. (b) Rank-2 CMVs occur in two possible topologies. If the eigenvalues' signs are equal, one obtains a ``disk" of either color (left) while if their signs  are opposite, one obtains a ``clover" (right). (c) Rank-3 CMVs occur in two possible topologies. If all eigenvalues' signs are equal, the CMV is a sphere; if one eigenvalue has a sign opposite the other two, one obtains a ``wheel-and-axle".}
	\label{class}
\end{figure*}

\section{Recipe for generating a correlation matrix visual (CMV)}\label{recipe}

We are interested in visualizing the two-point connected spin correlations. Accordingly, we define the one-spin and two-spin observables. The single spin Bloch vector is
\begin{equation}
\mathbf{b}_{i} = ( b^x_{i}, b^y_{i}, b^z_{i} ) = (\langle\sigma^x_{i} \rangle, \langle\sigma^y_{i} \rangle, \langle\sigma^z_{i}\rangle ) = \langle\boldsymbol\sigma_{i} \rangle 
\end{equation} 
where $i$ labels the particle and $\sigma^{x,y,z}$ are the Pauli operators. The correlations between a pair of spins $i$ and $j$ are 
\begin{equation}\label{smallc}
c_{ij} = 
\begin{pmatrix}
c^{xx}_{ij} & c^{xy}_{ij} & c^{xz}_{ij} \\
c^{yx}_{ij} & c^{yy}_{ij} & c^{yz}_{ij} \\
c^{zx}_{ij} & c^{zy}_{ij} & c^{zz}_{ij}
\end{pmatrix}
\end{equation}
where $c^{\mu\nu}_{ij}= \langle \sigma^{\mu}_i \sigma^{\nu}_j \rangle$ and $\mu, \nu \in\{ x,y,z\}$. The  connected correlations are
\begin{equation}\label{bigC}
C^{\mu\nu}_{ij} = c^{\mu\nu}_{ij} - b^{\mu}_ib^{\nu}_j  \ .
\end{equation}
Matrix elements $C^{\mu\nu}_{ij}$ are real for $i\neq j$. To obtain a geometric visualization of the correlation matrix $C_{ij}$, our first step is to relate it to the quadratic form
\begin{equation}\label{quad}
Q(C_{ij},\mathbf{r}) =  \mathbf{r}^{T} C_{ij} \mathbf{r} = \sum_{\mu,\nu \in \{x,y,z\}} C^{\mu\nu}_{ij} r^{\mu} r^{\nu}
\end{equation}
where $\mathbf{r} = (x,y,z)$. 
%In this work, we focus on the correlations between two spins that are part of a many-body system. 
One can visualize a quadratic form through its level sets (surfaces of constant value).
%, here known as quadric surfaces. 
This visualization has the drawback that the level sets of some quadratic forms are not compact (e.g. a hyperboloid) and thus not easily represented in a finite space. To associate a finite object with each quadratic form, we plot the level sets of
\begin{equation}\label{figquad}
Q_f(C_{ij},\mathbf{r})=\frac{Q(C_{ij},\mathbf{r})}{(1+r^2)^{3/2}}~ .
\end{equation} 
The choice of denominator is fairly arbitrary, and other choices, such as multiplying $Q$ by $e^{-r}$, could be made. We find that the choice in Eq.(\ref{figquad}) 
%gives a large dynamic range across which 
keeps the figure size roughly proportional to the size of correlations [App.~\ref{App3}]. We plot 
%one of these 
a level set (in red) where $Q_f$ attains a given positive value $P$, and another (in blue) where it attains $-P$. We typically we choose $P=0.01$.  We refer to this picture of level sets as a correlation matrix visualization (CMV). 

This method of visualization is sensitive only to the symmetric part of the correlation matrix. For such correlations, $Q_{ij}$ is in one-to-one correspondence with $C_{ij}$. Thus the CMVs are equivalent to the correlation matrices. Generalizations of our scheme to visualize $N$-body correlations are discussed in the Appendix~\ref{App0} and~\ref{App-irred}, and to visualize asymmetric correlation matrices in Appendix \ref{App2}. Although we work primarily with connected correlations $C_{ij}$, our scheme works just as well for correlations $c_{ij}$. 

Furthermore there is a one-to-one correspondence between the connected correlation matrix together with the Bloch vectors and the reduced two-spin density matrix  $\rho_{ij}$. This can be written in terms of the Pauli matrices as
\begin{equation}
\rho_{ij} = \frac{1}{4} \sum^3_{\alpha,\beta=0} S^{\alpha\beta}_{ij}~\sigma^{\alpha}_i \otimes \sigma^{\beta}_j 
\end{equation}
where $\sigma^{\alpha} \in \{1\!\!1,\sigma^x,\sigma^y,\sigma^z\}$ and $S^{\alpha\beta}_{ij} = \text{Tr}(\rho_{ij}~\sigma^{\alpha}_i \otimes \sigma^{\beta}_j )$ giving 
\begin{equation}
S_{ij} = 
\begin{pmatrix}
1 & b^x_i & b^y_i & b^z_i \\
b^x_j & c^{xx}_{ij} & c^{xy}_{ij} & c^{xz}_{ij} \\
b^y_j & c^{yx}_{ij} & c^{yy}_{ij} & c^{yz}_{ij} \\
b^z_j & c^{zx}_{ij} & c^{zy}_{ij} & c^{zz}_{ij}
\end{pmatrix} .
\end{equation}
Consequently, these three things are equivalent (still under the assumption that the correlation matrix is symmetric): (1) the CMV along with the Bloch vectors, (2) the correlation matrix $C_{ij}$ along with the Bloch vectors, and (3) the two-spin density matrix. We will see that examining the CMVs makes apparent physics that is hidden in the latter two representations.

\section{How to read a CMV}\label{read}

In this section, we discuss two essential aspects: how to read a CMV and how to  characterize it. One of the biggest advantages of our visualization is that one can deduce the magnitude of the correlations along any direction directly from the size of the CMV in that direction. The connected correlations between a pair of spins along a direction $\mathbf{e}$ is given as 
\begin{eqnarray}
&&\hspace{0.in}\langle (\boldsymbol{\sigma}_i \cdot \mathbf{e})  (\boldsymbol{\sigma}_j \cdot \mathbf{e} ) \rangle - \langle \boldsymbol{\sigma}_i \cdot \mathbf{e}\rangle  \langle \boldsymbol{\sigma}_j \cdot \mathbf{e} \rangle   \nonumber \\
&&\hspace{0.5in}{}= \sum_{\mu,\nu \in \{x,y,z\}}\langle (\sigma^{\mu}_ie^{\mu}) (\sigma^{\nu}_j e^{\nu})\rangle - \langle \sigma^{\mu}_i e^{\mu} \rangle \langle \sigma^{\nu}_j e^{\nu}\rangle  \nonumber \\
&&\hspace{0.5in}{}=\mathbf{e}^T C^{\mu\nu}_{ij} \mathbf{e} \nonumber \\
&&\hspace{0.5in}{}=Q(C_{ij},\mathbf{e}) \label{eq:correlation-along-e-CMV-size}
\end{eqnarray}
When $Q(C_{ij},\mathbf{e})$ is large in absolute value, the CMV (with the appropriate sign) will be large in this direction as well~(see Appendix~\ref{App3}). Consequently, one can read off the size of correlations along a certain direction by the size of the CMV in that direction. 

Using this geometric interpretation, we now proceed to interpret the CMVs in Fig.~\ref{class}. For example, consider the dumbbell-like CMV in Fig.~\ref{class}(a) lying along the $x$-axis. This CMV has its largest correlations along a line parallel to the x-axis which passes through its center. More specifically, $C^{xx}\neq0$ while $C^{yy}=C^{zz}=0$. Similarly consider the clover-shaped CMV in Fig.~\ref{class}(b). Visually one can infer that $C^{xx}, C^{yy} \neq 0$ but the correlation along a line 45$^{o}$ rotated from $x$ in $xy$ plane is zero. Finally the sphere in Fig.~\ref{class}(c) indicates that the correlation components are rotationally symmetric.

\begin{figure}[t]
	\centering
	\includegraphics[width=\columnwidth]{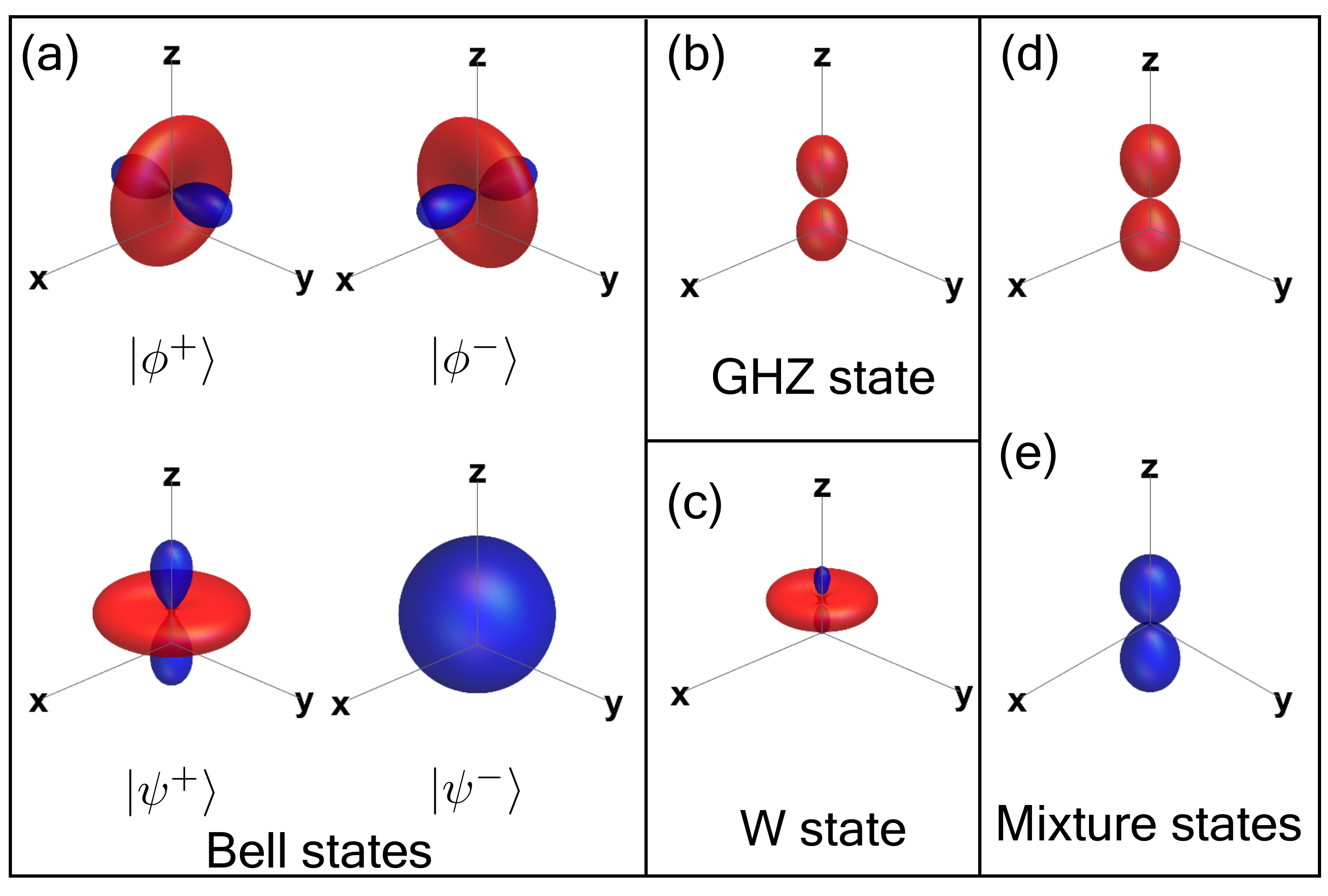}
	\caption{CMVs for prototypical states from density matrix, $\rho=\ket{\Psi}\!\bra{\Psi}$: (a) Bell states, $|\Psi\rangle =  |\phi^{\pm}\rangle = \left(|\downarrow\downarrow\rangle \pm |\uparrow\uparrow\rangle\right)/\sqrt{2}$, Bell states, $|\Psi\rangle =  |\psi^{\pm}\rangle =  \left(|\downarrow\uparrow\rangle \pm |\uparrow\downarrow\rangle\right)/\sqrt{2}$, (b) Two spin correlations  in a three spin GHZ state, $|\Psi\rangle = \left(|\downarrow\downarrow\downarrow\rangle + |\uparrow\uparrow\uparrow\rangle\right)/\sqrt{2}$, (c) Two spin correlations  in a three spin W state, $|\Psi\rangle = \left(|\downarrow\downarrow\uparrow\rangle + |\downarrow\uparrow\downarrow\rangle + |\uparrow\downarrow\downarrow\rangle\right)/\sqrt{3}$.  (d) CMVs for simple mixtures of product states: $\rho = \left(|\downarrow \downarrow\rangle\langle \downarrow \downarrow| + |\uparrow \uparrow \rangle\langle \uparrow \uparrow| \right)/2$ and (e) 
	$\rho = \left(|\downarrow \uparrow\rangle\langle \downarrow \uparrow| + |\uparrow \downarrow \rangle\langle \uparrow \downarrow| \right)/2$.}
	\label{bell}
\end{figure}

The shape and color of a CMV can be related to the eigenvalues of the correlation matrix. By shape we mean geometrical properties of the CMV that are invariant under rotation. Let us therefore consider a rotation matrix $R$ that rotates our original CMV such that $\mathbf{r}' = R \mathbf{r} $. Then
\begin{equation}
Q(C_{ij},\mathbf{r}') = \mathbf{r}^{T} (R^{T} C_{ij} R) \mathbf{r} = (R\mathbf{r})^T C_{ij} (R \mathbf{r})= Q(C'_{ij},\mathbf{r})
\end{equation}
where we define $C_{ij}'=R^T C_{ij} R$. In other words, two CMVs that are related by a rotation $R$ are associated with correlation matrices $C_1$ and $C_2$ that are similar to each other under $R$ (and thus have the same eigenvalues). The converse is also true. That is, if we have two correlation matrices that are related by rotations, then they are represented by CMVs that have the same shape, but  different orientations. As an example, the Bell states $(\phi^\pm, \psi^-)$, which are shown in Fig.~\ref{bell}(a), illustrate the physical significance of this rotational equivalence. Each of these CMVs can be rotated to any other because these three Bell states are equivalent under global rotations. 

Although the topology and colors are the most visually apparent features of the CMV, they do not uniquely describe a state. Other aspects such as the  size of the CMV and the corresponding Bloch vectors are also needed. For example, the CMV of the GHZ state and the mixed states in Fig.~\ref{bell}(b) and (d) respectively both have the dumbbell topology but differ in size. The same is true when we compare the three Bell states ($ \phi^{\pm}, \psi^{-}$) with the W state. Note that the states chosen in Fig.~\ref{bell} have Bloch vectors equal to zero.  In general this is not the case, as we will see in the subsequent sections.

\section{CMVs for physically realizable spin models}\label{real}  

Here we calculate and describe the CMVs that arise in prototypical many-body systems. We focus on a few examples that are motivated by experiments on ultracold matter. In particular, we consider the non-equilibrium dynamics in the Ising model evolved from a product state ferromagnet (Secs.~\ref{Ising1} and~\ref{Ising2}), the dynamics of the Fermi-Hubbard model (Sec.~\ref{fermi}) evolved from a product state canted antiferromagnet, and the equilibrium correlations of the transverse field Ising model across its phase diagram (Sec.~\ref{trans}). All of these models have been realized in ongoing experiments, as briefly described in their respective sections. Conveniently, they also  can be solved exactly. We focus on one-dimensional systems with nearest-neighbor (NN) couplings for simplicity. In all cases we find that although the correlations can be complicated, the CMVs behave in a simple manner. 

\subsection{Coherent Ising model}\label{Ising1}

The Ising model has been extensively studied in the context of ferromagnetism and phase transitions \cite{Sachdev}, and has been applied to study cooperative behavior in fields as far removed as biology \cite{Baake}. In ultracold matter, there are many proposals and realizations of Ising Hamiltonians for example using trapped ions \cite{Islam583,Kim, Blatt, Garttner}, dipolar molecules \cite{Gorshkov, Alexey, Manmana, Yan, Barnett, Micheli, Kaden0, Kaden2} and atoms \cite{Paz, Paz2}, Rydberg atoms \cite{Zeiher, Takei, Weimer1, Robert, Weimer2, Labuhn, Schauss1455,Mukherjee,Gil} and tilted Bose-Hubbard systems \cite{Sachdev2,Meinert1259}. Ref.~\cite{Hazzard} overviews  these systems and others in which the Ising model can and has been realized.

We consider the dynamics of the one-dimensional nearest-neighbor Ising model without a transverse field or decoherence. We initiate the dynamics from a product state. This is described by the Hamiltonian
\begin{equation}\label{}
H_{\rm I} = - J\sum_{i} \sigma^z_i \sigma^z_{i+1}
\end{equation}
where $J$ is the interaction term. We study the dynamics of this model initiated from the product state $\ket{\theta \theta \theta\cdots}$ where
\begin{equation}
\ket{\theta}=\cos \theta/2 \ket{\uparrow}+ \sin \theta/2 \ket{\downarrow}.\label{eq:theta}
\end{equation}
With no loss of generality, we have assumed that the spins are initially in the $x$-$z$ plane. One experimentally straightforward way to produce this initial state is to rotate each spin, after it is prepared in the single-spin ground state $\ket{\downarrow}$, by using a global laser pulse. This is the standard first step of any Ramsey experiment \cite{Zeiher,Knap}. The system is then allowed to evolve from the initial state under $H_{\rm I}$.

Over time, correlations build up in the system. Refs.~\cite{Kastner, Hazzard, Mauritz} calculate the single-spin expectations and two-spin correlations. The single-spin observables are
\begin{align}
b^z_k(t) =\langle \sigma^z_k(t) \rangle &= \cos \theta \\\label{sigma_z}
b^x_k(t) =\langle \sigma^x_k(t)\rangle &= \text{Re}\left(\sin\theta\left[g^{+}(Jt)\right]^2 \right) \\
b^y_k(t) =\langle \sigma^y_k(t)\rangle &= \text{Im}\left(\sin\theta\left[g^{+}(Jt)\right]^2 \right)
\end{align}
where 
\begin{equation}\label{g}
g^{\pm}(x) = \cos^2 (\theta/2)~e^{-i2x} \pm \sin^2 (\theta/2)~e^{i2x}
\end{equation}
 The two point correlations are
\begin{align}
c^{xx}_{jk}(t) &= \frac{1}{4} \left(c^{++}_{jk}+c^{--}_{jk}+c^{+-}_{jk}+c^{-+}_{jk} \right) \label{cohcorr1}\\
c^{yy}_{jk}(t) &= \frac{1}{4} \left(c^{+-}_{jk}+c^{-+}_{jk}-c^{++}_{jk}-c^{--}_{jk} \right) \label{cohcorryy} \\
c^{zz}_{j,k}(t) &= \expec{\sigma^z_j(t)}\expec{\sigma^z_k(t)} \label{cohcorrzz}\\
c^{xy}_{jk}(t) &= \frac{1}{4i} \left(c^{++}_{jk}-c^{--}_{jk}-c^{+-}_{jk}+C^{-+}_{jk} \right) \label{cohcorrxy} \\
c^{xz}_{jk}(t) &= \frac{1}{2} \left(c^{+z}_{jk} + c^{-z}_{jk}  \right) \label{cohcorrxz} \\
c^{yz}_{jk}(t) &= \frac{1}{2i} \left(c^{+z}_{jk} - c^{-z}_{jk}  \right) \label{cohcorr2}
\end{align}
where 
\begin{eqnarray}
c^{+\pm}_{k,k+n}(t)
&=& \left\{
\begin{array}{ll}\label{Ising_c1}
\sin^2\theta~\left[ g^{+} (Jt) g^{+} (\pm Jt)\right] ~&\text{for $n=1$.}\\ \\
\sin^2\theta~g^{+}(Jt\pm Jt)\\
\times \left[g^{+}(Jt))g^{+}(\pm Jt))\right] ~&\text{for $n=2$.} \\ \nonumber\\
\sin^2\theta~\left[g^{+}(Jt)g^{+}(\pm Jt)\right]^2 ~&\text{for $n>2$.}
\end{array}
\right.\\ \\
c^{+z}_{k,k+n}(t) &=& 
\label{Ising_c2}
\begin{cases}
\sin \theta~ g^{-}(Jt) g^{+} (Jt) ~&\text{for $n=1$}   \\
\sin \theta~\left[g^{+} (Jt) \right]^2 ~&\text{for $n>1$}
\end{cases},
\end{eqnarray}
and the other correlations can be calculated from the identities
\begin{eqnarray}
c^{-\pm}_{k,k+n}(t) &=& \left(c^{+\mp}_{k,k+n}(t)\right)^{*} \label{eq:adjoint-identities}\\
% c^{--}_{k,k+n}(t) &=& \left(c^{++}_{k,k+n}(t)\right)^{*}  \\
c^{-z}_{k,k+n}(t) &=& \left(c_{k,k+n}^{+z}(t)\right)^*. \label{eq:adjoint-identities2}
\end{eqnarray}

\begin{figure}[t]
	\centering
	\includegraphics[width=\columnwidth]{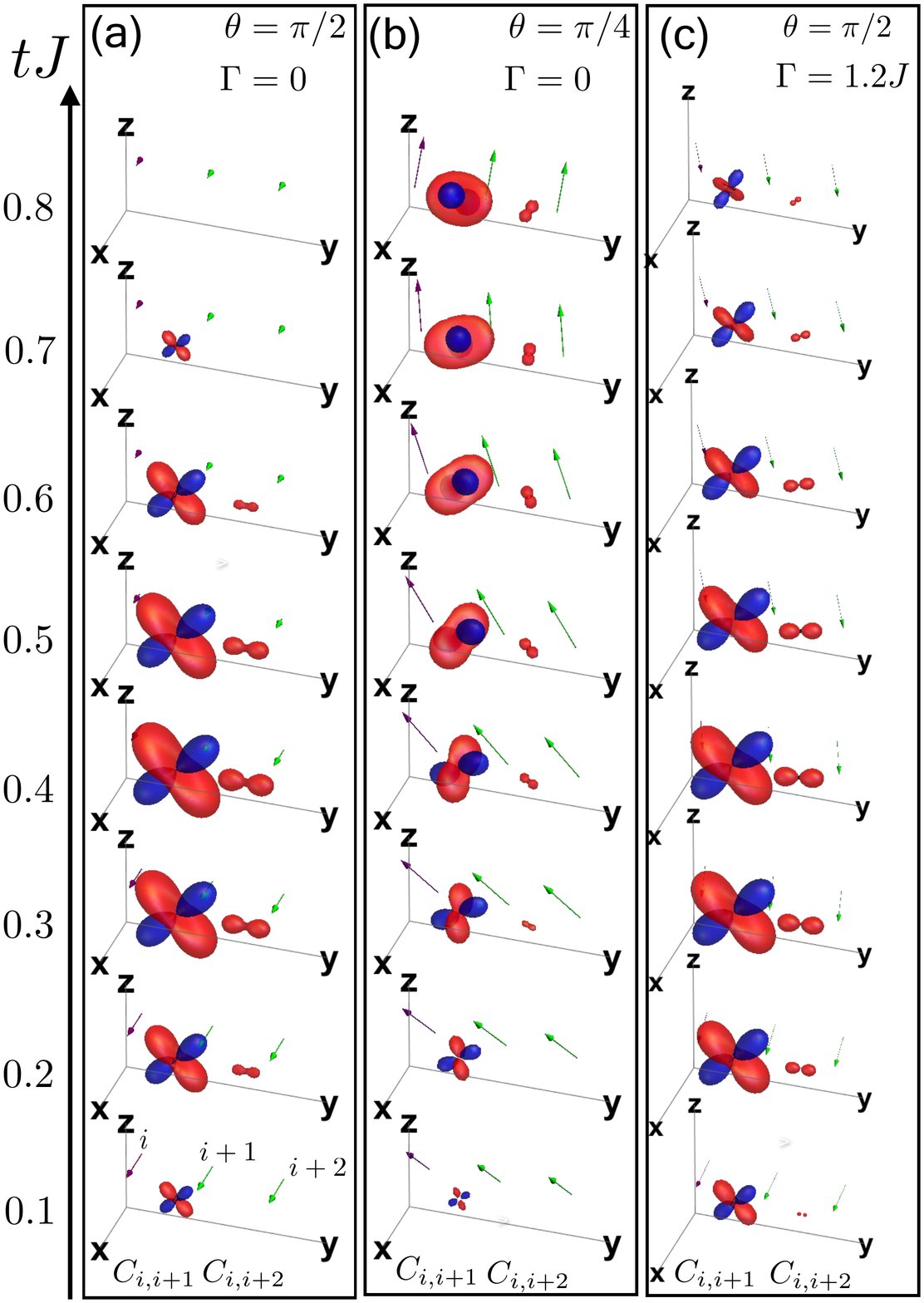}
	\caption{ Dynamics of the nearest-neighbor one-dimensional Ising model (no transverse field) initiated from a state with all the spins aligned. Arrows (green and purple) indicate Bloch vectors of three adjacent spins belonging to an infinite spin chain. At every instant, the correlations are depicted with respect to the first spin (purple). (a)  Coherent Ising dynamics (no dissipation) for $\theta=\pi/2$ initial state. (b)  Ising dynamics for $\theta=\pi/4$ initial state. (c) Ising dynamics with dissipation for same $\theta$ as (a). The CMVs for (b) and (c) have been magnified by factors of 1.2 and 1.5 respectively to be comparable to (a) for convenient visualization.}
	\label{Ising}
\end{figure}	
Fig.~\ref{Ising}(a) shows the coherent dynamics for the initial state $\theta=\pi/2$. The NN correlations are described by a clover oriented with its petals pointing at 45$^{o}$ in the yz-plane. It grows and shrinks periodically. At its maximum size, the clover fattens to have some wheel-and-axle character. 

Fig.~\ref{Ising}(b) shows the same dynamics for the initial state $\theta=\pi/4$. (This is the same example as that used in Fig.~\ref{setup}). The dynamics is qualitatively similar to the $\pi/2$ case, except (i) the initial CMV is rotated to orient the clover perpendicular to the initial Bloch vector, (ii) the clover is slightly smaller  (note that the size of the CMVs have been magnified for easier visualization in Fig.~\ref{Ising}(b).), and (iii) a precession about the $z$ axis is superimposed.  This precession is a mean field effect in which every spin experiences a local field along the $z$-axis due to the interactions with the background spins. Hence the CMVs precess along with the Bloch vectors.

As expected for a model with short-range interactions, next-nearest-neighbor (NNN) correlations are smaller than the NN correlations. More interestingly, the structure of the NNN CMV is qualitatively different. It is a simple dumbbell that grows and shrinks. In the Supplemental material we have movies for the coherent dynamics for longer times \cite{supp}.

This dynamics is exemplary of the utility of the CMVs. Remarkably, although the components of the correlations evolve in an extremely complicated manner, as shown in Fig.~\ref{setup}(b-c), the CMVs behave simply: a growing and shrinking clover superimposed on a rotation. If one were to consider the dynamics of the many components $C_{ij}^{\alpha\beta}$, it would be difficult to describe them, remember them, or form a mental picture of them. In contrast, the CMVs provide a description of the dynamics in terms of growth and rotation of figures that is simple to describe, remember, and visualize.

\subsection{Ising model with decoherence}\label{Ising2}

There are deep questions about the fate of correlations in open quantum systems that the CMVs may be useful in illuminating. Furthermore, any system is inevitably coupled, however weakly, to some environment, so understanding the effects of this decoherence on the correlations is an important practical goal.

With this in mind, we consider the dynamics of the Ising model in the presence of an incoherent spontaneous emission. We choose this form of decoherence both due to its simplicity and  because it is an important decoherence mechanism in many experiments. The  system's reduced density matrix satisfies a master equation with Markovian dissipation,
\begin{equation}
\dot{\rho} = -i\left[H_{\rm I},\rho  \right] + \mathcal{L}(\rho)
\end{equation}
where $\mathcal{L}(\rho)$ is the Lindblad term
\begin{equation}
\mathcal{L}(\rho) = \frac{\Gamma}{2}\sum_j \left(2\sigma^{-}_j\rho\sigma^{+}_j - \sigma^{+}_j\sigma^{-}_j\rho - \rho\sigma^{+}_j\sigma^{-}_j   \right).
\end{equation} 
From Refs. \cite{Foss-Feig, Foss-Feig2}, the single-spin expectations are
\begin{align}
b^z_k(t) = \langle \sigma^z_k(t) \rangle &= (e^{-\Gamma t}-1)+ e^{-\Gamma t} \cos \theta \\
b^x_k(t) = \langle \sigma^x_k(t)\rangle &= \text{Re}\left(e^{-\Gamma t} \Phi^2(J,t) \right) \\
b^y_k(t) = \langle \sigma^y_k(t)\rangle &= \text{Im}\left(e^{-\Gamma t} \Phi^2(J,t) \right)
\end{align}
where
\begin{equation}
\Phi(J,t) = e^{-\Gamma t/2}\left[ \cos(st) + \frac{\Gamma t}{2}~\text{sinc}(st) \right]
\end{equation}
with $\operatorname{sinc}(x)=\sin(x)/x$ and
\begin{equation}
s=2\left(i\frac{\Gamma}{4}-J\right).
\end{equation}
The correlations $c^{\alpha\beta}_{jk}$ can be calculated using 
Eqs.~(\ref{cohcorr1}-\ref{cohcorr2}) and Eqs.~(\ref{eq:adjoint-identities}-\ref{eq:adjoint-identities2}), where for the dynamics with decoherence
\begin{align}
c^{+\pm}_{k,k+n}(t)
    &= \begin{cases}
        e^{-2\Gamma t}\Phi(J,t)\Phi(\pm J,t) & \text{for $n=1$}\\ 
        e^{-2\Gamma t} \Phi(J\pm J,t) \Phi( J,t)\Phi(\pm J,t)  &\text{for $n=2$}\\ 
        e^{-2\Gamma t}\left[ \Phi(J,t)\Phi(\pm J,t)\right]^2 &\text{for $n>2$}  
       \end{cases} \\
c^{+z}_{k,k+n}(t)
     &= \begin{cases}
          e^{-\Gamma t}\Psi( J,t)~\Phi( J,t) &\text{for $n=1$} \\
          e^{-\Gamma t}~\Phi^2( J,t) &\text{for $n>1$}
\end{cases}
\end{align}
where 
\begin{equation}
\Psi(J,t) = e^{-\Gamma t/2}~(is-\frac{\Gamma}{2})t~\text{sinc}(st)~.\\
\end{equation}

Figure~\ref{Ising}(c) depicts the CMV for the Ising dynamics with decoherence for $\theta=\pi/2$ (see also the movie in the Supplemental material \cite{supp}). The shape for the NN correlations is a clover throughout the dynamics. It never obtains the substantial wheel-and-axle character that it did for the coherent case. 

Despite the similarities, adding decoherence qualitatively modifies the $\theta=\pi/2$ CMV dynamics in two ways. The first, trivial, effect is that eventually the CMV vanishes. This happens because the correlations vanish when all of the spins have damped to the ground state (pointing along the negative $z$-axis). 

Second, the CMV precesses  (rotates about the $z$-axis), which becomes visually apparent starting around $tJ\sim 0.7$.
Unlike the coherent $\theta=\pi/4$ dynamics, the rate of precession of the CMV (and Bloch vectors) with decoherence grows with time. In fact, it is proportional to the $z$ component of the Bloch vector, as this is essentially a mean field effect similar to that occurring for $\theta=\pi/4$. 

\subsection{Fermi-Hubbard model}\label{fermi}

In this section, we study the dynamics of correlations of the Fermi-Hubbard model after a sudden quench to the non-interacting limit, which is similar to various quenches considered in  Refs.~\cite{Vekua, Dao, Gluza, PhysRevLett.117.190602, WhiteHuletHazzard}. This can be accomplished by sweeping across a Feshbach resonance or sufficiently lowering the lattice depth. The equilibrium properties of the Fermi-Hubbard model are well studied in cold atoms, and studying its quench dynamics is an exciting frontier~\cite{Barmettler,Greif, Mathy, Duarte, Hulet, Hofrichter, Taie, Strohmaier, simon}. Great progress has been made in terms of imaging fermions \cite{Zimmermann,Sanner,Torben}, achieving Mott states \cite{Jordens,Parsons,Greif3} and generating short-range AFM correlations \cite{cheuk:observation_2016,Greif4,Greif3,Greif2,Greif,Hulet,Boll}. With these techniques, experimentally preparing the states that we study and measuring the resulting correlations has become feasible.

\begin{figure}[t]
	\centering
	\includegraphics[width=\columnwidth]{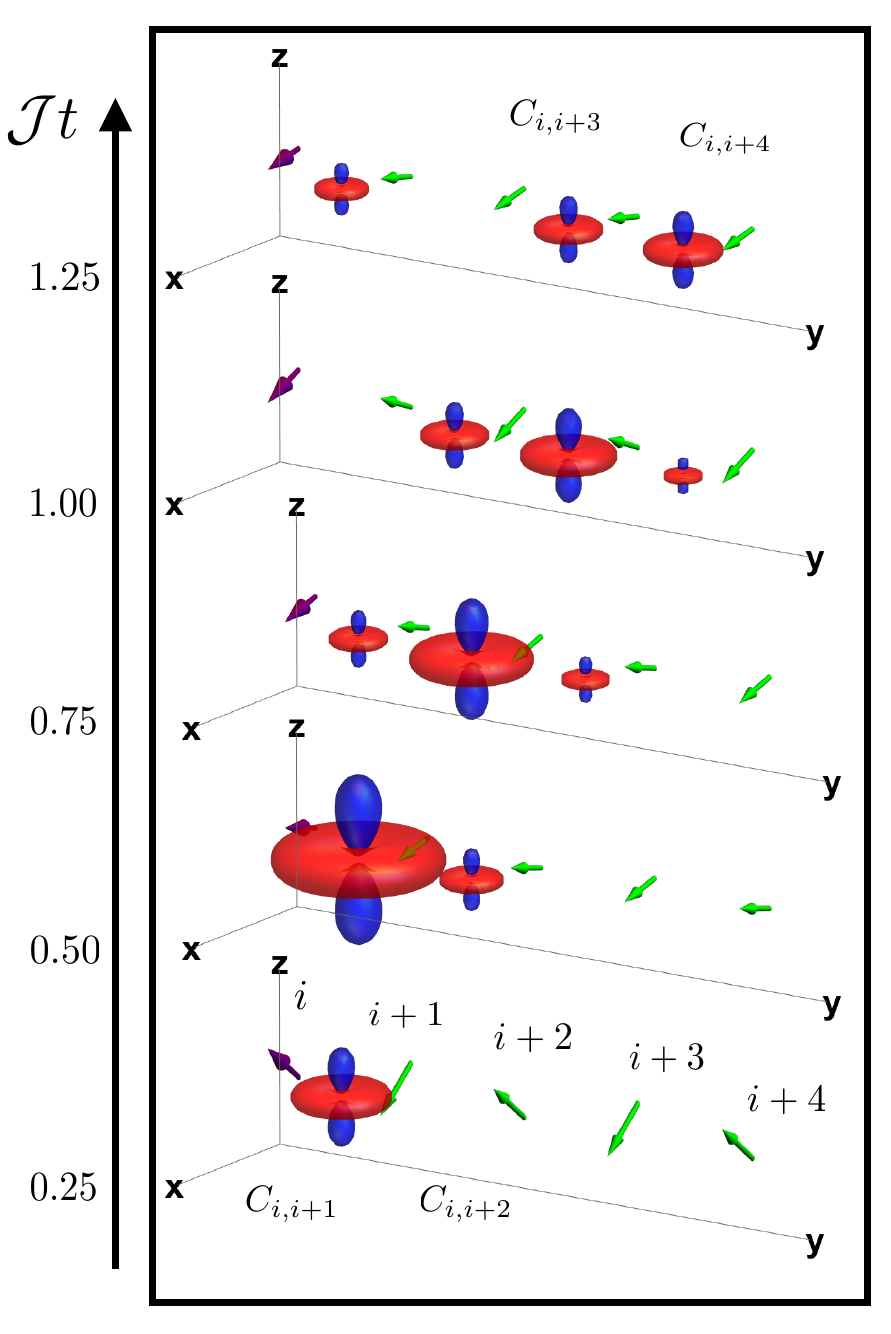}
	\caption{ Dynamics of non-interacting spin-1/2 fermions initiated from an initial one-spin-per-site (Mott insulating) staggered spin product state is shown. Time advances vertically from bottom to top with each panel corresponding to the evolution of the correlation at a given time.}
	\label{FH}
\end{figure}	

Although our calculations allow general product states as initial conditions, here we concentrate on an initial canted antiferromagnetic product state that is discussed below.  As in prior sections, we concentrate on the one dimensional case for simplicity. The Fermi-Hubbard Hamiltonian is 
\begin{equation}
H = - \mathcal{J} \sum_{\langle i,j \rangle, \sigma} c^{\dagger}_{i\sigma} c_{j\sigma}  + U \sum_i n _{i\uparrow} n _{i\downarrow} 
\end{equation}
where $\mathcal{J}$ is tunneling, $U$ is the on-site interaction and $\sigma = \uparrow$, $\downarrow$. The $c^{\dagger}_{i\sigma}$($c_{i\sigma}$) are fermionic creation (annihilation) operators and $n _{i\sigma} = c^{\dagger}_{i\sigma}c_{i\sigma}$ is the number operator for spin $\sigma$ at site $i$. We calculate the correlations for $U=0$ between sites $q$ and $r$,
\begin{align}
\langle \sigma_{q}^{\mu}(t)\sigma_{r}^{\nu}(t)\rangle =& \sum_{abcd}\sigma_{ab}^{\mu}\sigma_{cd}^{\nu}\langle c_{qa}^{\dagger}(t)c_{qb}(t)c_{rc}^{\dagger}(t)c_{rd}(t)\rangle \nonumber \\
=& \sum_{ijkl}A_{qi}^{*}(t)A_{qj}(t)A_{rk}^{*}(t)A_{rl}(t) \nonumber \\
&\times\sum_{abcd}\sigma_{ab}^{\mu}\sigma_{cd}^{\nu}\langle c_{ia}^{\dagger}(0)c_{jb}(0)c_{kc}^{\dagger}(0)c_{ld}(0)\rangle
\end{align}
where $a,b,c,d \in \{ \uparrow, \downarrow\}$. The $A_{jl}$ are the propagators
\begin{eqnarray}
A_{jl} &=& \frac{1}{N}\underset{k}{\sum}\exp\left[ik(j-l)+i E_{k}t\right]  \nonumber \\
&=& (-i)^{|j-l|} J_{|j-l|}(2\mathcal{J} t).
\end{eqnarray}
Here $J_{m}\left(z\right)$ is a Bessel function of the first kind and
\begin{equation}
E_{k} =-2\mathcal{J} \cos\left(k\right)
\end{equation}
is the dispersion relation where we have assumed unit lattice spacing. 

Because the initial state is a product state, the expectation value of a many-body operator can be factored into a product of single site expectation values.  
For an arbitrary product state, we find that the Bloch vector are
\begin{equation}
b_{i}^{\mu}(t)=\sum_{ab}\sigma_{ab}^{\mu}\sum_j |A_{ij}(t)|^2f_{j}\left(ab\right)\
\end{equation}
and the correlation functions are
%\begin{widetext}
\begin{eqnarray}
&&  c^{\mu\nu}_{qr}(t) = \sum_{abcd} \sigma_{ab}^{\mu} \sigma_{cd}^{\nu}  \Big\{\sum_{i}|A_{qi}(t)|^{2} |A_{ri}(t)|^{2} 
%\right. 
\nonumber \\ 
&&\hspace{0.3in}{}\times 
\Big[g_i(abcd) -f_i(ab) f_i(cd) -f_i(ad)\left(\delta_{bc}-f_i(cb) \right) \Big] \nonumber \\ 
&& \hspace{0.1in}
%\left.  
{}+\Big(\sum_{i} |A_{qi}(t)|^{2} f_i(ab)\Big)\Big(\sum_{k}|A_{rk}(t)|^{2}  f_k(cd)\Big)  %\right.
\nonumber \\ 
&&  \hspace{0.1in}{} 
%\left. 
+\Big(\sum_{i}A_{qi}^{*}(t)A_{ri}(t) f_i(ac) \Big)
%\right.    
\nonumber \\
&& \hspace{0.2in}
 {}\times
	\Big( \sum_{j} A_{qj}(t)A_{rj}^{*}(t) \left(\delta_{bc}-f_j(cb)\right) \Big)\Big\} 
\end{eqnarray}
where we have defined
\begin{align}
f_{j}\left(ab\right) &=\left\langle c_{ja}^{\dagger}c_{jb}^{\phantom{\dagger}}\right\rangle \\
g_{j}\left(abcd\right) &=\left\langle c_{ja}^{\dagger}c_{jb}^{\phantom{\dagger}}c_{jc}^{\dagger}c_{jd}^{\phantom{\dagger}}\right\rangle 
\end{align}
with the expectation values taken at time $t=0$.

As an example, consider the product state
\begin{align}\label{fermiini}
\left|\psi\right\rangle &=\underset{k}{\bigotimes}\begin{cases}
\left|\nearrow\right\rangle _{k} & k\,\textrm{even}\\
\left|\searrow\right\rangle _{k} & k\,\textrm{odd}\end{cases}
\end{align}
where  $\ket{\nearrow}=\cos(\pi/8)\ket{\downarrow}+\sin(\pi/8)\ket{\uparrow}$ and $\ket{\searrow}=\cos(3\pi/8)\ket{\downarrow}+\sin(3\pi/8)\ket{\uparrow}$ (refer to Eq.~(\ref{eq:theta})). 
Similar initial states can be prepared experimentally using  magnetic field gradients  \cite{Schrader}, polarization field gradients in optical superlattices \cite{Lee, Sebby}, or spin-changing collisions in double wells \cite{Trotzky295,Chen,Gerbier,Widera}.
For this initial state
%, Eq.~\eqref{fermiini}, 
we have 
\begin{align}
f_{j}\left(ab\right)  &=
\left[\frac{1}{2}+\frac{\left(-1\right)^{j+\delta_{a\downarrow}}}{2\sqrt{2}}\right] \delta_{ab} + \frac{1-\delta_{ab}}{2\sqrt{2}} \\
g_{j}\left(abcd\right) &=
\left[\frac{1}{2}+\frac{\left(-1\right)^{j+\delta_{a\downarrow}}}{2\sqrt{2}}\right] \delta_{bc}\delta_{ad} +\frac{\delta_{bc}(1-\delta_{ad})}{2\sqrt{2}}.
\end{align}

Figure~\ref{FH} depicts the dynamics of a system that starts with a canted antiferromagnet as its initial state and then is quenched to a non-interacting system (see also the movie in the Supplemental material \cite{supp}). We find that the maximum correlation with respect to the first spin (purple) initially shifts over time from NN to NNN. At longer times, the influence of correlations continues to spread over wider spatial regions.  The propagation of correlations in a lattice is commonly referred as the light cone \cite{Lieb1972,Bravyi,Bravyi, Cheneau_light,Richerme, Jurcevic}. It should also be noted that the maximum size any CMV can attain decreases with distance and time.  
Interestingly, the CMVs have the wheel-and-axle shape. This shape is invariant under rotations about the $z$ axis, which indicates an emergent symmetry in the correlations that is absent from the initial state.

\begin{figure*}[t]
	\centering
	\includegraphics[scale=0.5]{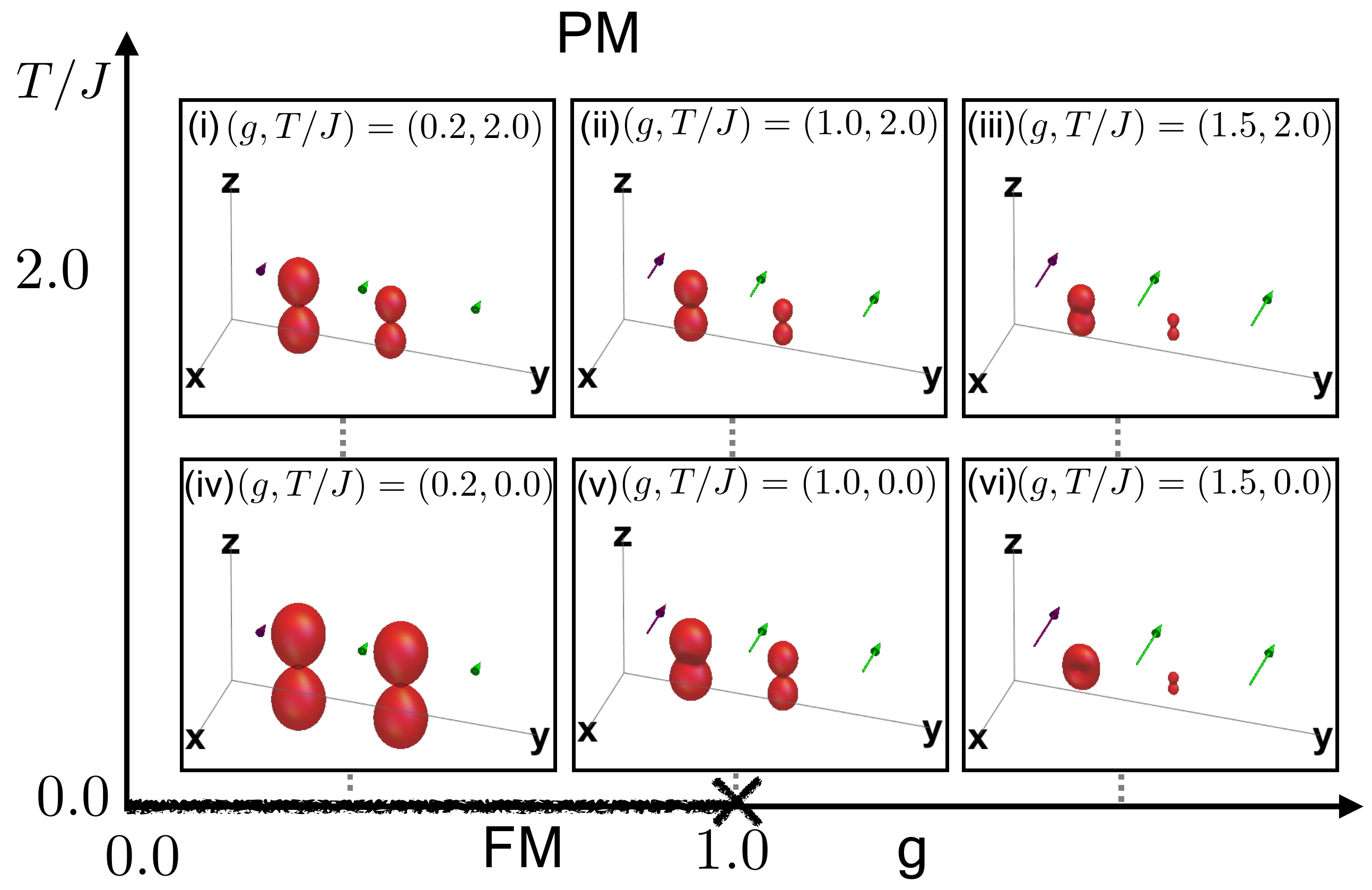}
	\caption{Correlations are shown throughout the equilibrium phase diagram for a one dimensional NN transverse field Ising model where FM stands for ferromagnetic phase and PM for the paramagnetic phase. Each panel in (i)-(vi) contains the CMVs corresponding to a specific value for the parameters $(g,T/J)$. Left and right CMVs in each panel are NN and NNN correlations respectively.}
	\label{Ising_trans}
\end{figure*}

\subsection{Transverse Ising phase diagram}\label{trans}

In contrast to the previous sections, which involved non-equilibrium dynamics,  here we focus on equilibrium correlations. Specifically, we study the one-dimensional transverse Ising model with NN interactions,
\begin{equation}
H_{\rm T} = H_{\rm I} - h \sum_i \sigma^x_i,
\end{equation}
where $h>0$ is the transverse field. We define  $g=h/J$. With this definition, $g_c=1$ is the critical value of $g$ that separates the ferromagnetic and paramagnetic phases at zero temperature. 

The single spin expectations are \cite{PFEUTY}
\begin{eqnarray}
\expec{\sigma^x} &=& 0 \\
\expec{\sigma^y} &=& 0 \\
\expec{\sigma^z} &=& D_0,
\end{eqnarray}
and the  correlations are \cite{PFEUTY}
\begin{eqnarray}	
C^{xx}_{i,i+n} 
&=& 	
\begin{vmatrix}
D_{0} & D_{n} \\
D_{-n} & D_{0}
\end{vmatrix} 
- D^2_{0} \nonumber \\
&=& -D_{-n}D_{n} \\
\nonumber \\
\nonumber  \\
%\end{eqnarray}
%\begin{eqnarray}
C^{yy}_{i,i+n}
&=& 	
\begin{vmatrix}
D_{-1} & D_{0} & \hdots & D_{n-2} \\
D_{-2} & D_{-1} & \hdots & D_{n-3} \\
\vdots & \vdots & \ddots & \vdots \\
D_{-n} & D_{-n+1} & \hdots & D_{-1} 
\end{vmatrix} \\
%\end{eqnarray}
%\begin{eqnarray}
\nonumber \\
\nonumber  \\
C^{zz}_{i,i+n} 
&=& 	-
\begin{vmatrix}
D_{1} & D_{2} & \hdots & D_{n} \\
D_{0} & D_{1} & \hdots & D_{n-1} \\
\vdots & \vdots & \ddots & \vdots \\
D_{-n+2} & D_{-n+3} & \hdots & D_{1} 
\end{vmatrix} \\
\nonumber \\
\nonumber  \\
C^{xy}_{i,j} &=& C^{yz}_{i,j} = C^{xz}_{i,j} = 0
\end{eqnarray}
where
\begin{equation}
D_{n}
= -\int_{-\pi}^{\pi} \frac{dk}{2\pi} \left[1 - 2v^2_k +2iu_kv_k\right] e^{ikn} \tanh (\frac{\omega_k}{T})~.
\end{equation}
Here $T$ is the temperature, we have set Boltzmann constant's $k_B=1$, $\omega_k$ is given as
\begin{equation}
\omega_k = 2J\sqrt{1+g^2-2g\cos k},
\end{equation}	
and $u_k$ and $v_k$ are given as
\begin{eqnarray}
u_k &=& \frac{2J\sin k}{\sqrt{2\omega_k (\omega_k - 2J(g-\cos k))}} \\ 
v_k &=& \frac{\omega_k - 2J(g-\cos k)}{\sqrt{2\omega_k (\omega_k - 2J(g-\cos k))}}~. 
\end{eqnarray}
Fig.~\ref{Ising_trans} shows CMVs in different regions of the equilibrium phase diagram of the transverse Ising model. At $g=0$, we have a dumbbell shaped CMV for all values of $T$. The dumbbell shape and its evolution with $g$ and $T$ is of some physical interest. To obtain some insight into this, note that in the (purely classical) limit of a Hamiltonian depending only on the $\sigma^z_i$, the density matrix is $\sum_{qr \in \{\uparrow \downarrow\}} \rho_{qr} \ket{qr}\!\bra{qr}$. This has correlations along $z$ and vanishing $x$ and $y$ correlations, leading to a dumbbell shape. Upon perturbing this system with a transverse field, quantum fluctuations create coherences and correlations involving the $x$ and $y$ directions. This is reminiscent of the  fact that in the similar steering ellipsoid visualization of Ref.~\cite{Garon}, a single non-zero eigenvalue is insufficient to obtain entanglement.

On increasing the transverse field $g$, there are two significant trends. First the Bloch vectors grow along the $x$-direction as expected. Secondly, the CMVs gradually change from a dumbbell to a disc like shape [see Fig.~\ref{Ising_trans} (vi)], which indicates that there are non-zero correlations in the  $yz$ plane. This is due to   quantum fluctuations. 

For extremely large $g$, a new symmetry  emerges corresponding to rotations about the $x$ axis. One way to understand this is by the mapping presented in Ref.~\cite{Richerme} showing that this model for $g\rightarrow \infty$ maps to an XX model with this symmetry. However, we note that by examining the correlations with CMVs, the emergent symmetry is immediately apparent even without prior knowledge of  this mapping. 

Increasing temperature leads to a dumbbell shape in all cases. This is clearest in Fig.~\ref{Ising_trans}(iii). This occurs as the thermal fluctuations overwhelm the quantum fluctuations, even though the correlations remain strong.

The NNN correlations show similar behavior to NN correlation for small $g$. The main difference is that the NNN correlations are smaller in magnitude. However the NNN correlations only obtain their emergent symmetry at much larger $g$, and they do so by vanishing in this limit. 

\section{Conclusion}

We generate a three dimensional geometric object by associating spin correlation matrices with a quadratic form. We apply this visualization scheme to various prototypical correlated quantum states, and to a variety of ultracold systems. We demonstrate that phenomena that look complicated and mysterious when analyzed by the components of their correlations become simple and intuitive when described geometrically. For example, the CMVs describing correlations often pulsate and rotate in fairly simple ways.

The simplicity uncovered by our work opens up exciting questions: Are there ways to determine qualitatively the shapes that appear in the CMVs, especially for models that are not exactly solvable? Are there general rules that underly the simple motions of the CMVs? For example, can we understand under what circumstances CMVs grow and shrink, or when they simply rotate? It will also be interesting to uncover how a given manipulation of the CMV shape can be engineered via Hamiltonian evolution or other quantum operations. Using this insight to control the state would have applications in quantum information and many-body physics. It would be intriguing to combine this with the methods in Ref.~\cite{Carr2}, where each edge on their graph would be associated with a CMV between the corresponding pair of spins.

Two-spin CMVs can be generalized in various ways.
One can associate three- and higher-spin correlations with geometric objects [see Appendix~\ref{App0}]. The number of correlation components grows exponentially with the number of spins involved, $N_s$,  specifically there are $3^{N_s}$ components. Thus the correlations may have a complexity far exceeding two spin correlations. It would be especially interesting if equally simple motions underlie these higher-spin correlations.

KRAH acknowledges the Aspen Center for Physics for its hospitality while part of this work was performed. JH acknowledges the SCI REU program for giving him the opportunity to do this work at Rice university. This work was supported with funds from the Welch foundation, Grant No. C-1872. We also acknowledge discussions with Charles Xu and Kenneth Wang.

\bibliography{corrvisual}

\newpage

%\section*{Appendix}

\appendix

\section{Visualizing asymmetric correlation matrix}\label{App2}

Our method of using quadratic forms depends only on the symmetric 
part of the correlation matrix. To visualize an arbitrary correlation matrix $C_{ij}$, note that it can be split into a sum of symmetric 
and antisymmetric parts, $\mathcal{S}_{ij}^{\alpha\beta}=(1/2)(C_{ij}^{\alpha\beta}+C_{ij}^{\beta\alpha})$ and $\mathcal{A}_{ij}^{\alpha\beta}=(1/2)(C_{ij}^{\alpha\beta}-C_{ij}^{\beta\alpha})$, respectively. 
The former may be visualized as in the main text, while the antisymmetric piece can be identified as usual with a pseudovector $(a,b,c)$. The psuedovector can be visualized as an arrow (similar to the Bloch vectors) or by level sets of linear forms 
${L}(\vec{r})=a x + b y + cz$ (perhaps adding an appropriate denominator to render them compact), much as we use level 
sets of ${Q}(\vec{r})$ for the symmetric piece. Another method to handle asymmetric correlations is to take singular value decompositions of the relevant matrix and consider left and right eigenvectors~\cite{Gamel}.

\section{Generalization to many-spin correlations}\label{App0}

Let the connected correlation matrix for $N$ spin-1/2's be the rank-$N$ tensor $C_{ij \hdots N}^{\mu\nu\hdots\gamma}=\expec{\sigma^\mu_i \sigma^\nu_j \cdots \sigma^\gamma_N }-\,\cdots$ where ``$\cdots$" indicates terms to be subtracted in order to produce the desired connected correlation. One can define a CMV  by
\begin{equation}
F(C_{i j \hdots N}, \mathbf{r}) = \sum_{\mu,\nu \hdots \gamma\in\{x,y,z\}} C^{\mu\nu \hdots \gamma}_{i j \hdots N} r^{\mu} r^{\nu} \hdots r^{\gamma}.
\end{equation}
Similar to the two-spin case, the extent of the geometrical object in a direction $\alpha$ measures the size of the  correlations in that direction. Also similar to the two spin case, this is now sensitive only to the totally permutationally symmetric component of the correlations.

\section{Irreducible components of $N$-body correlations \label{App-irred}}

In order to gain insight into the correlations, it can be useful to decompose the tensor $C_{ij\hdots N}$ into its spherically irreducible components, a method employed in Ref.~\cite{Garon}. This method makes more explicit the properties of the figures under global spin rotations and includes both symmetric and asymmetric correlations in a uniform manner.  
We decompose
\begin{equation}
C_{ij\hdots N} = \sum_{l\in L} \sum^l_{m = -l} a_{lm}(C_{ij\hdots N}) T_{lm}
\end{equation} 
where $L=\{0,1,\ldots,N\}$, $T_{lm}$ are the components of the spherical tensor $T_{l}$ with rank $l$, and $a_{lm}$ are the coefficients of the expansion, which characterize $C_{ij\hdots N}$. To visualize $C_{ij\hdots N}$, one can plot  
\begin{equation}
f(C_{ij\hdots N};\theta,\phi)  = \sum_{l\in L} \sum^l_{m = -l} a_{lm}(C_{ij\hdots N}) Y_{lm}(\theta,\phi)\label{eq:irred-decomp-Ylm}
\end{equation}
where $Y_{lm} (\theta,\phi)$ are the spherical harmonics. If one prefers a three dimension image, similar to the CMVs, one can add a dependence on $r$ and plot level sets; e.g. one can plot level sets of $f(C_{ij\hdots N};\theta,\phi)/(1+r^2)^{3/2}$. 

A convenient extension to this is to consider associate each irreducible sector with its own geometric object. Thus, one associates the $l=0$ terms with one object, the $l=1$ terms with another, and so on. To do this, one restricts the sum in Eq.~\eqref{eq:irred-decomp-Ylm} to the $l$ corresponding to the object.  This representation is convenient because under global spin rotations, objects of shapes in the rank-$l$ irreducible space transform only among each other, and not into shapes in other spaces. 

The case of two spin-1/2's provides a familiar example. The correlation tensor can be decomposed into a sum of $l=0,1,2$ components $C = C_{0}+ C_{1}+C_{2}$ (suppressing $ij\hdots N$ indices for brevity) with
\begin{align}
C_{0}^{\mu\nu}  &= \delta_{\mu\nu}\frac{\operatorname{Tr}(C)}{3} \\
C_{1}^{\mu\nu}  &= \frac{C^{\mu\nu}-C^{\nu\mu}}{2} \\
C_{2}^{\mu\nu}  &= \frac{C^{\mu\nu}+C^{\nu\mu}}{2} - C_{0}.
\end{align}  
To compare this decomposition with the method used in the main text, note that there we plot the entire symmetric part of $C$, i.e. $C_{2}^{\mu\nu} +C_{0}^{\mu\nu}$. This more directly allows one to read off the size of the total correlations in a given direction, but at the cost of a slightly more complicated transformation under rotations.

\section{Relating the size of CMV to the magnitude of correlations}\label{App3}

\begin{figure}[t]
	\centering
	\includegraphics[width=\columnwidth]{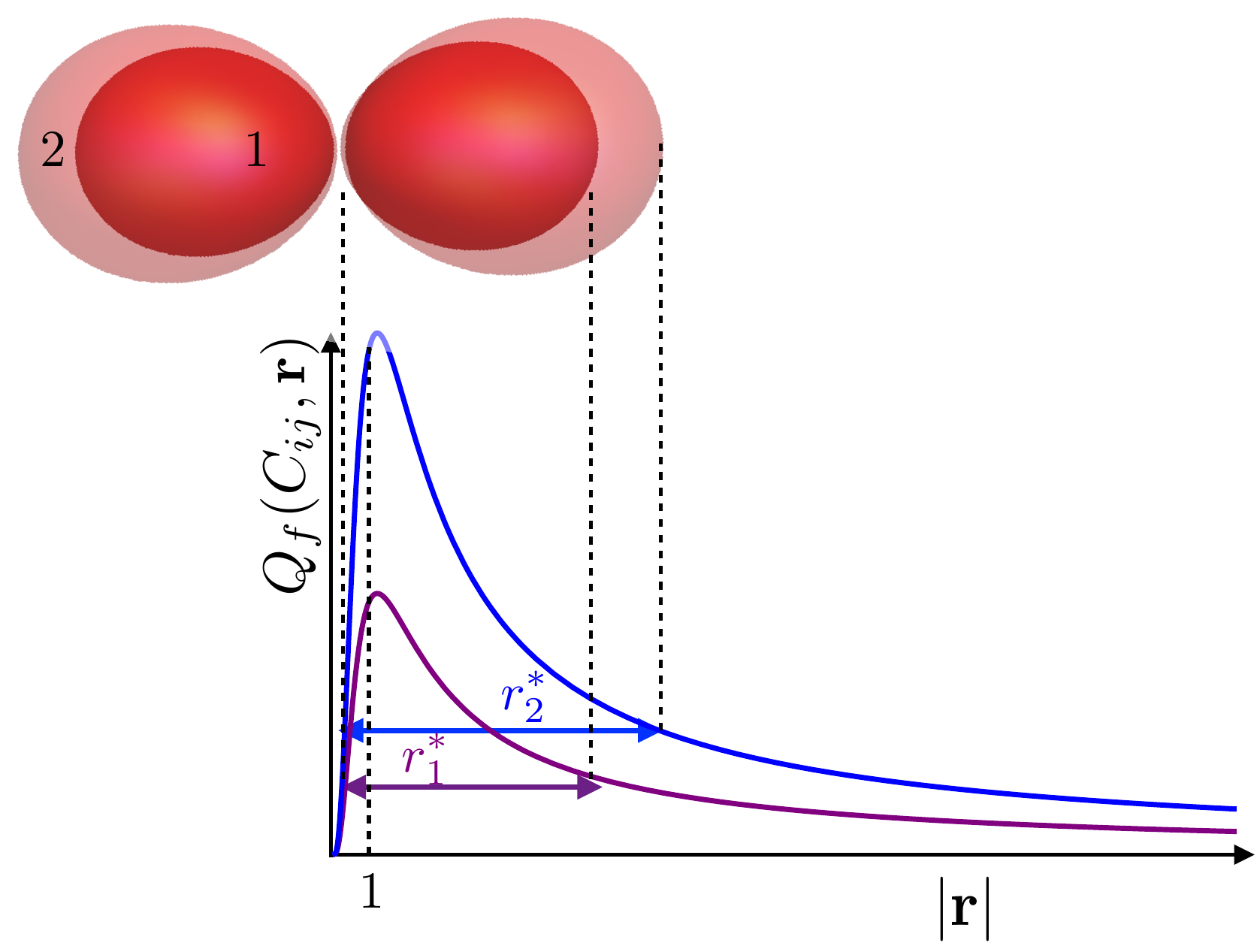}
	\caption{Two CMVs are shown corresponding to level sets of $Q_f({C}_{ij}=0.1,r)=0.1$ (labeled as 1)  and $Q_f({C}_{ij}=0.2,r)=0.1$ (labeled as 2). The size of each lobe belonging to CMV 1,2 is given by $r^{*}_{1,2}$, as suggested by Eq.~\eqref{size}. }
	\label{Appendix}
\end{figure}

In the main text, Sec.~\ref{read} discussed how to read information off of a CMV, and in particular it claimed that the extent of the 
CMV along a unit direction $\mathbf{\hat e}$ indicates the size of ``connected correlations in that direction", 
 $ {\mathcal C}_{ij}({\mathbf{\hat e}})
=\langle 
 (\boldsymbol{\sigma}_i \cdot \mathbf{\hat e})  (\boldsymbol{\sigma}_j \cdot \mathbf{\hat e} ) \rangle 
 - \langle \boldsymbol{\sigma}_i \cdot \mathbf{\hat e}\rangle  \langle \boldsymbol{\sigma}_j \cdot \mathbf{\hat e} \rangle $. 
The main step towards this connection is Eq.~\eqref{eq:correlation-along-e-CMV-size}, which established that 
$ {\mathcal C}_{ij}(\mathbf{\hat e})$ is 
equal to $Q(C_{ij},\mathbf{\hat e})$. 
Then, the interpretation of the CMV is validated by noting that the value of  $Q(C_{ij},\mathbf{\hat e})$ determines the extent of the CMV along $\mathbf{\hat e}$. 

The purpose of this Appendix is to establish this intuitive link between $Q(C_{ij},\mathbf{\hat e})$  and the size of the CMV along $\mathbf{\hat e}$, and to give more insight into it. Fig.~\ref{Appendix} demonstrates the key idea: as $Q(C_{ij},\mathbf{\hat e})$ is increased, the graph of $Q_f$ as a function of the distance along that direction is increased. Due to peaked shape of this function, the inner and outer parts of the level set get farther apart, increasing the size of the shape in that direction.

To establish this geometric fact algebraically, we will calculate the separation of the inner and outer level sets where $Q(C_{ij},r^*)=P$, $r^*_{\rm in}$ and $r^*_{\rm out}$ respectively. We will assume that the inner level set $r^*_{\rm in}\ll 1$ and $r^*_{\rm out} \gg 1$, not because we expect this to be valid generally, but because this case is illustrative of the connection that we seek to establish and the math is particularly simple. 
Equation~(\ref{figquad}) can be rewritten as
\begin{align}
Q_f &= \frac{\langle (\mathbf{\sigma}_i \cdot \mathbf{r})(\mathbf{\sigma}_j \cdot \mathbf{r})\rangle - \langle  \mathbf{\sigma}_i \cdot \mathbf{r}\rangle \langle \mathbf{\sigma}_j \cdot \mathbf{r}\rangle}{(1+r^2)^{3/2}} \nonumber \\
&= \frac{r^2\langle (\mathbf{\sigma}_i \cdot \hat{r})(\mathbf{\sigma}_j \cdot \hat{r})\rangle - \langle  \mathbf{\sigma}_i \cdot \hat{r}\rangle \langle \mathbf{\sigma}_j \cdot \hat{r}\rangle}{(1+r^2)^{3/2}} \nonumber \\
&= \frac{r^2}{(1+r^2)^{3/2}}{\mathcal C}_{ij}(\mathbf{ \hat e}).
\end{align}
 In the limit considered,  $r^{*}_{\rm out } = {\mathcal C}_{ij}(\mathbf{\hat e})/P$ and $r^{*}_{\rm in } = \sqrt{P/{\mathcal C}_{ij}(\mathbf{\hat e})}$. Thus the size of a CMV along the direction $\mathbf{r}$ is given as 
\begin{equation}\label{size}
r^*_{\rm size} = r^{*}_{\rm out } - r^{*}_{\rm in} = \left({\mathcal C}_{ij}(\mathbf{\hat e})/P - \sqrt{P/{\mathcal C}_{ij}(\mathbf{\hat e})}\right)~.
\end{equation}
From the above expression, one sees that as the
magnitude of ${\mathcal C}_{ij}(\mathbf{\hat e})$ increases, so does the size of the CMV along that direction. This interpretation still holds without making the asymptotic assumptions we have here, but the CMV size along $\mathbf{\hat e}$ is no longer related to the value of $Q(C_{ij},\mathbf{\hat e})$ by a simple proportionality.

\end{document}